\documentclass[twocolumn,showpacs,prc]{revtex4}
\usepackage{mathrsfs}
\usepackage{amsmath, txfonts}
\usepackage{graphicx,,booktabs,bm}
\usepackage{longtable,lscape}
\usepackage{overpic,multirow}
\usepackage{color}
\usepackage{amssymb}

\def\thalf{{\textstyle{3\over 2}}}

\usepackage{indentfirst}
\begin{document}
\title{The internal structures of  the nucleon resonances $N(1875)$ and $N(2120)$}
\author{Jun He}\email{junhe@impcas.ac.cn}
\affiliation{
Theoretical Physics Division, Institute of Modern Physics, Chinese Academy of Sciences,
Lanzhou 730000, China
}
\affiliation{
Research Center for Hadron and CSR Physics,
Lanzhou University and Institute of Modern Physics of CAS, Lanzhou 730000, China
}
\affiliation{
State Key Laboratory of Theoretical Physics, Institute of
Theoretical Physics, Chinese Academy of Sciences, Beijing  100190,China
}

\begin{abstract}

A nucleon resonance with spin-parity $J^P=3/2^-$ and mass about 2.1
GeV is essential to reproduce the photoproduction cross sections for $\Lambda(1520)$
released by the LEPS and CLAS Collaborations. It can be explained as the third nucleon resonance state
$[3/2^-]_3$ in the constituent quark model so that there is no position to
settle the $N(1875)$ which is listed in the PDG as the third $N3/2^-$ nucleon
resonance. An interpretation is proposed that the
$N(1875)$ is from the interaction of a decuplet baryon $\Sigma(1385)$
and a octet meson $K$, which is favored  by a calculation of binding energy and decay pattern in a Bethe-Salpeter
approach.

\end{abstract}

\maketitle
\maketitle


\section{introduction}

In new versions of the Review of Particle physics (PDG) after the year
2012~\cite{Agashe:2014kda}, there are four $N3/2^-$ states, $N(1520)$,
$N(1700)$, $N(1875)$ and $N(2120)$.  The two-star state
$N(2080)$ in previous versions has been split into a three-star
$N(1875)$ and a two-star $N(2120)$ based on the evidence from BnGa
analysis~\cite{Anisovich:2011fc}.

Usually the $N(1520)$ and the $N(1700)$ are assigned to states with orbital
angular momentum $L=1$ in quark model, and mixing effect is very
important to explain the decay pattern of these states~\cite{He:2003vi}. The situation
for the internal structures of two $N3/2^-$ states with higher mass,
the $N(1875)$ and the $N(2120)$, is much less unclear. In quark model,
the $N(1875)$ and the $N(2120)$ are in the mass region of $N=3$ band states of which the masses and
decay patterns were predicted~\cite{Capstick:1998uh,Capstick:1992uc}. However, the explicit
correspondence between predicted and observed  states is
unclear. In Large $N_c$ QCD, the third and fourth $N3/2^-$ states have masses $2101\pm14$ and $2170\pm42$ MeV, respectively \cite{Matagne:2014lla}. Klempt and others claimed that the $N(1875)$ is the missing third $N(3/2^-)$
state in mass region $1800-1900$ MeV with orbit angular momentum $L=1$ and
radial excitation number ${\rm N}=1$
~\cite{Klempt:2009pi}, which is also supported by the Ads/QCD~\cite{Brodsky:2014yha}.
Their conclusion is only based on a comparison between predicted and observed masses.
As enlightened by  Isgur,  ``in a complex system like the
baryon resonances, predicting the spectrum of states is not a very
stringent test of a model''~\cite{Isgur:1999jv}.  Decay pattern provides more information
about hadron internal structure.

Many analyses suggested that a $N3/2^-$ state with mass about 2.1 GeV
is essential to explain experimental results~\cite{Kohri:2009xe,Kim:2011rm,Xie:2010yk,He:2012ud}. Before the year 2012,
it is related to the only $N3/2^-$ state listed in the PDG with mass higher than 1.8
GeV, the $N(2080)$, and explained as the third state $[N3/2^-]_3$ predicted
in the constituent quark model. For example, the $N(2080)$ is found to play the most
important role in the photoproduction of $\Lambda(1520)$ off proton
target~\cite{Xie:2010yk,He:2012ud}.
Recently, the CLAS Collaboration at Jefferson National Accelerator
Facility released their exclusive photoproduction cross section for
the $\Lambda$(1520) for energies
from near threshold up to a center of mass energy $W$ of 2.85 GeV with
large range of the $K$ production angle~\cite{Moriya:2013hwg}.
The reanalyses about the new data in Refs.~\cite{He:2014gga,Xie:2013mua}
confirmed the previous conclusion that a nucleon resonance near 2.1 GeV, $N(2120)$, is essential to
reproduce the experimental data~\cite{Xie:2010yk,He:2012ud}.

\section{Role of the $N(2120)$ in the $\Lambda(1520)$ photoproduction}

In the following it will be shown why the $N(2120)$ should be assigned
to the third state $[N3/2^-]_3$ in the constituent quark model in line with the theoretical framework in Ref.~\cite{He:2014gga}. There are five $N3/2^-$ states in $N=3$ band, of which the radiative and $\Lambda(1520)K$ decay amplitudes were predicted in Refs.~\cite{Capstick:1998uh,Capstick:1992uc} as listed in Table.~\ref{Tab: Resonances}.

\begin{table}[h!]
\renewcommand\tabcolsep{0.36cm}
\renewcommand{\arraystretch}{1.}
\caption{The $N3/2^-$  nucleon resonances and their decay amplitudes predicted in the relativistic quark model~\cite{Capstick:1998uh,Capstick:1992uc}. The mass $m_R$, helicity
	amplitudes $A_{1/2,3/2}$ and partial wave decay amplitudes
	$G(\ell)$ are in the unit of MeV, $10^{-3}/\sqrt{\rm{GeV}}$ and
	$\sqrt{\rm{MeV}}$, respectively.}
\begin{tabular}{c|rrrrr}
	\toprule[1.pt] State  & $m_R$ & $A^p_{1/2}$ &  $A^p_{3/2}$ &
  $G(\ell_1)$ &  $G(\ell_2)$  \\\hline
 $[N\textstyle{3/2}^-]_3$ &1960 &   36  & -43 & $-2.6 $ & $-0.2$
  \\
 $[N\textstyle{3/2}^-]_4$ & 2055 &   16  &   0 & $-0.5$ & $0.0$
  \\
 $[N\textstyle{3/2}^-]_5$ & 2095 &   -9  & -14 & $0.4$ & $0.0$
  \\
   $[N\textstyle{3/2}^-]_6$ & 2165 &  $--$   &  $--$  & $0.4$ & $0.0$
  \\
 $[N\textstyle{3/2}^-]_7$ & 2180 &   $--$  & $--$ & $1.1$ & $0.1$
  \\
 \bottomrule[1.pt]
\end{tabular}
\label{Tab: Resonances}
\end{table}

The predicted  radiative and strong decay amplitudes suggest the
importance of the nucleon resonance $[N3/2^-]_3$ in the $\Lambda(1520)$ photoproduction, which is the first state in $N=3$ band states
and the third state in all nucleon resonances with $J^P=3/2^-$ predicted in the constituent
quark model.

\begin{figure}[h!]
\begin{flushleft}
\includegraphics[width=0.45\textwidth]{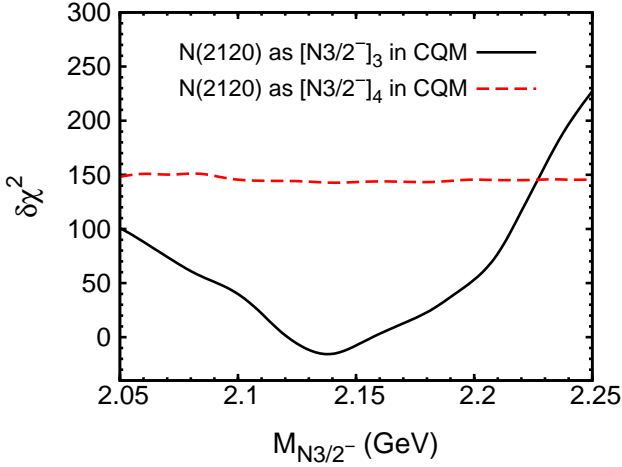}
\caption{
(Color online)
The change of $\chi^2$ in mass scan. The
solid and dashed lines are for the results with assuming the $N(2120)$ as state$[N3/2^-]_3$ and with assuming the $N(2120)$ as state $[N3/2^-]_4$, respectively.
\label{Fig: chi2} }
\end{flushleft}
\end{figure}

In Ref.~\cite{He:2014gga}, based on the high precision experimental data released by the
CLAS and LEPS Collaborations recently, the interaction mechanism of
the photoproduction of $\Lambda$(1520) off a proton target is investigated
within a Regge-plus-resonance approach. The inclusion of the $N(2120)$ as state
$[N3/2^-]_3$ in the constituent quark model reduced the $\chi^2$ obviously.
In that work, mass and width are fixed at 2.12 GeV and 0.33 GeV,
respectively. Here, a mass scan is made for the $N(2120)$ by fitting the data from the CLAS and LEPS Collaborations. Except mass, the width $\Gamma_R$, which was fixed at 0.33 GeV
in previous work~\cite{He:2014gga}, is also set as a free parameter. The behavior of $\chi^2$ is presented in Fig.~\ref{Fig: chi2}.

Here the results with assuming the $N(2120)$ as state $[N3/2^-]_3$
and with assuming the $N(2120)$ as $[N3/2^-]_4$ are provided.
If the $N(2120)$ is assumed to be the third state
$[N3/2^-]_3$ in the constituent quark model, the change of $\chi^2$ will decrease and reach minimum at 2.13 GeV with the increase of mass. If assumed to be
the fourth state $[N3/2^-]_4$, the change of $\chi^2$ keep stable around
150, which means that the experimental data can not be well reproduced.
Obviously, the $N(2120)$ should be assigned as state $[N3/2^-]_3$ instead of
state $[N3/2^-]_4$ in the constituent quark model.  Since the $N(1875)$
is much lower than the $N(2120)$, it is unnatural to assign it to the
fourth or higher states. The first and
second states in the constituent quark model have been assigned to
the four-star $N(1520)$ and the three-star $N(1700)$ in the PDG, which has been confirmed by many experimental and theoretical evidences~\cite{Agashe:2014kda}. Hence,
there is no position to settle the $N(1875)$ in the constituent quark
model.

The state $[N3/2^-]_3$ predicted in the constituent quark model is much
lower than the $N(2120)$ even if model uncertainty, about 100 MeV, is considered. It is well-known that
loop effect will lead to mass shift. The state $[N3/2^-]_3$ has a large decay
width in $\Lambda(1520) K$ channel as predicted in the constituent quark model~\cite{Capstick:1998uh,Capstick:1992uc}. Moreover, the $\Lambda(1520)K$ threshold
is near bare mass of state $[N3/2^-]_3$. Hence,
the large difference between bare mass and observed mass can be explained by both  uncertainty of the constituent quark model and mass shift arising from the $\Lambda(1520) K$ loop effect.

\section{The $N(1875)$ as a $\Sigma(1385)K$ bound state}

Now that the $N(1875)$ can not be explained in the conventional quark model, it may be a exotic hadron.  In meson sector, some of particles which can not be explained in the quark
model framework, such as $XYZ$ particles observed in recent years, have been suggested to be
hadronic molecular states. The light scalars $a_0(980)$, $f_0(980)$
and $f_0(500)$ are often considered as meson-meson resonances. In
baryon sector, some authors proposed that the $\Lambda (1405)$ may be explained as a $N\bar{K}$ bound
state~\cite{Dalitz:1960du,Jido:2010ag,Hall:2014uca}. The mass of the $N(1875)$ is close to the $\Sigma(1385)K$ threshold, which encourages us to interpret $N(1875)$ as
a bound state of $\Sigma^*$ and $K$ (here and hereafter I denote
$\Sigma(1385)$ as $\Sigma^*$).  As said above, the internal
structure of a hadron can not be judged only through its mass. In this
work, both mass and decay pattern of $N(1875)$ as a
bound state of $\Sigma^*$ and $K$ will be calculated with a method developed based
on the covariant spectator formalism of the Bethe-Salpeter
equation~\cite{Gross:2010qm,Gross:2014wqa,Biernat:2013aka,Gross:2012sj,He:2014,He:2014nxa}, which has been used to study
the $B\bar{B}^*$ and the $D\bar{D}^*$ systems.

Analogous to Ref.~\cite{He:2014,He:2014nxa}, with help of onshellness of the heavy
constituent 1, $\Sigma^*$, the numerator of propagator $P_1^{\mu\nu}$ is rewritten as $\sum_{\lambda}u_{1\lambda}^\mu \bar{u}_{1\lambda}^\nu$ with $u^\mu_{1\lambda}$ being
the Rarita-Schwinger spinor with helicity $\lambda$. The equation for vertex
is in a form
\begin{eqnarray}
	&&|{\Gamma}_{\lambda}\rangle
	=\sum_{\lambda'}{\cal{V}}_{\lambda\lambda'}~{G}_0~
	|{\Gamma}_{\lambda'}\rangle,
\end{eqnarray}
with
$|{\Gamma}_{\lambda}\rangle=\bar{u}_\lambda^\mu|{\Gamma}_{\mu\mu'}\rangle
u_R^{\mu'}$
and ${\cal{V}}_{\lambda\lambda'}=\bar{u}_\lambda^\mu{{\cal
V}}_{\mu\nu'}u_{\lambda'}^{\nu'}$.
The rest of propagator $G_0$ for particle 1
and 2 with mass $m_1$ and $m_2$ written down in
the center of mass frame where $P=(W,{\bm 0})$  is
\begin{eqnarray}
G_0=2\pi i\frac{\delta^+(k_1^2-m_1^2)}{k_2^2-m_2^2}=2\pi
i\frac{\delta^+(k^0_1-E_1({\bm k}))}{2E_1({\bm k})[(W-E_1({\bm k})^2-E_2^2({\bm k})]},
\end{eqnarray}
where $k_1=(k_1^0,\bm
k)=(E_1({\bm k}),\bm k)$, $k_2=(k_2^0,-\bm
k)=(W-E_1({\bm k}),-\bm k)$ with $E_{1,2}({\bm k})=\sqrt{m_{1,2}^2+|\bm k|^2}$.

The integral equation  can be written explicitly as
\begin{eqnarray}
&&(W-E_1({\bm k})-E_2({\bm k}))\phi_{\lambda}({\bm k})\nonumber\\&=&\sum_{\lambda'}\int\frac{d{\bm k}'}{(2\pi)^3}V_{\lambda\lambda'}({\bm k},{\bm k}',W)\phi_{\lambda'}({\bm k}'),
\end{eqnarray}
with
\begin{eqnarray}
V_{\lambda\lambda'}({\bm k},{\bm k}',W)=\frac{i~\bar{\cal
	V}_{\lambda\lambda'}({\bm k},{\bm k}',W)}{\sqrt{2E_1({\bm
	k})2E_2({\bm k})2E'_1({\bm k}')2E'_2({\bm k}')}}, \label{Eq:
	Lp}
\end{eqnarray}
where the reduced potential kernel ${\cal
\bar{V}}_{\lambda\lambda'}=F({\bm k}){\cal{V}}_{\lambda\lambda'}F({\bm k}')$
with a factor as
$F({\bm k})=\sqrt{2E_2({\bm k})/( W-E_1({\bm k})+E_2({\bm k}))}$. The normalized wave function can be related to vertex as
$|\phi\rangle=N|\psi\rangle=NF^{-1}G_0~|{\Gamma}_{\lambda}\rangle$ with
the normalization factor $N({\bm k})=\sqrt{2E_1({\bm k})E_2({\bm k})/(2\pi)^5W}$.

Since $K$ is a pseudoscalar particle, it is forbidden to exchange pseudoscalar meson between $K$ and $\Sigma^*$.  The vector meson exchanges, $\rho$, $\omega$ and $\phi$, is dominant in the interaction. The potential kernel ${\cal V}$ can be obtained from the effective
Lagrangians describing the
interactions for vector mesons $V$ with $K$  and $\Sigma^*$,
\begin{eqnarray}
	{\cal L}_{KKV}&=&ig_{KK V} K^\dag
	{V}^\mu\partial_\mu K,\\
	{\cal L}_{\Sigma^*\Sigma^*V}&=&g_{\Sigma^*\Sigma^*V}
	\Sigma^{*\dag}_\mu[\gamma^\nu-\frac{\kappa_{\Sigma^*\Sigma^*V}}{2m_{\Sigma}}
	\sigma^{\mu\rho}\partial_\rho]{V}^\nu\Sigma_\mu.
\end{eqnarray}
In this work the isospin structures are following the standard form in
Ref.~\cite{deSwart:1963gc} and omitted in the Lagrangians. The
coupling constants for vector mesons $\rho$, $\omega$ and $\phi$
interacted with $K$ and $\Sigma^*$ can be obtained from $g_{\rho
\pi\pi}=6.199$ and $f_{\rho\Delta\Delta}=-4.30$ in quark
model~\cite{Matsuyama:2006rp} and relations
$g_{KK\rho}=g_{\rho\pi\pi}/2=g_{KK\omega}=\sqrt{2}g_{KK\phi}=g_{\rho\pi\pi}/2$ and
$g_{\Sigma^*\Sigma^*\rho}=-g_{\Sigma^*\Sigma^*\omega}=g_{\Sigma^*\Sigma^*\phi}/\sqrt{2}=g_{\Delta\Delta\rho}$
under $SU(3)$ symmetry. Here different definitions between Ref.~\cite{Matsuyama:2006rp} and this work have been
considered. Since the constituent 2 is off shell, a monopole form factor is introduced at the vertex for each off-shell kaon meson with mass $m_K$  as
$h(k^2)={\Lambda^4}/{[(m_{K}^2-k^2)^2+\Lambda^4]}$. The form factor for
the exchanged meson with mass $m_V$ is chosen as
$f(q^2)={(\Lambda^2-m_V^2)}/{(\Lambda^2-q^2)}$. Empirically the cut off $\Lambda$ should be not far from 1 GeV.

The 3-dimensional equation can be reduced to a one-dimensional equation with
partial wave expansion. The wave function has an angular dependent as
\begin{eqnarray}
&&\phi_{\lambda}({\bm
k})=\sqrt{\frac{2J+1}{4\pi}}D^{J*}_{\lambda_R,\lambda}(\phi,\theta,0)\phi_{\lambda,\lambda_R}(|{\bm
k}|),
\end{eqnarray}
where $D^{J*}_{\lambda_R,\lambda}(\phi,\theta,0)$ is the rotation matrix with $\lambda_R$ being the helicity of bound state with angular momentum $J$.
The potential after partial wave expansion is
\begin{eqnarray}
V_{\lambda\lambda'}^J(|{\bm k}|,|{\bm k}|')=2\pi\int
d\cos\theta_{k,k'}
d^{J}_{\lambda,\lambda'}(\theta_{k,k'})V_{\lambda\lambda'}({\bm
k},{\bm k}'),
\end{eqnarray}
where $\theta_{k,k'}$ is angle between ${\bm k}$ and ${\bm k}'$.
The one-dimensional integral equation reads
\begin{eqnarray}
	&&(W-E_1(|\bm k|)-E_2(|\bm k|))\phi^J_{\lambda}(|{\bm
	k}|)\nonumber\\&=&\sum_{\lambda'}\int  \frac{|{\bm k}'|^2d|{\bm
	k}'|}{(2\pi)^3}V^J_{\lambda\lambda'}(|{\bm k}'|,|{\bm
	k}'|)\phi^J_{\lambda'}(|{\bm k}'|).\label{Eq: final equation}
\end{eqnarray}

To study the decay property of a bound state, the information about coupling of a bound state to its constituents is essential. In literatures it is often achieved with the method proposed by Weinberg~\cite{Weinberg:1962hj,Faessler:2007gv}. In this work, the
vertex wave function, which contains the information about coupling of bound state to its constituents, is obtained during solving the binding energy. It make  a study of the decay pattern of the $\Sigma^*K$ bound state possible.

\begin{figure}[h!]
\begin{center}
\includegraphics[width=0.21\textwidth]{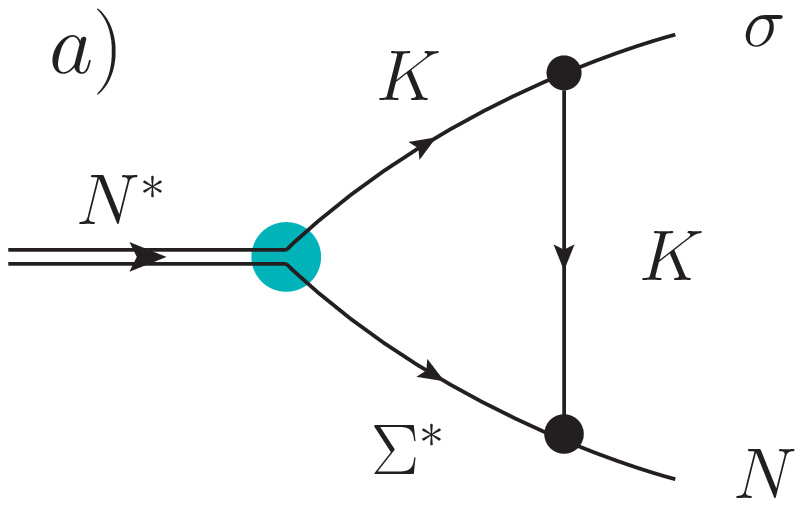}
\includegraphics[width=0.21\textwidth]{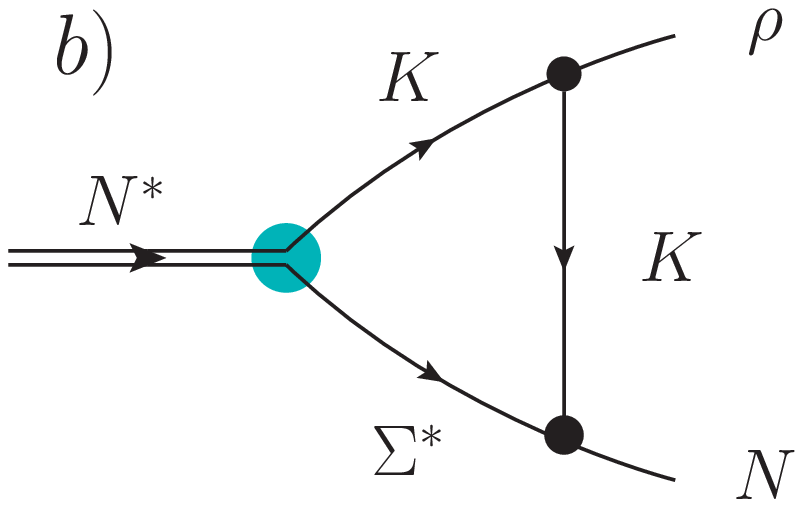}\\ \ \\
\includegraphics[width=0.21\textwidth]{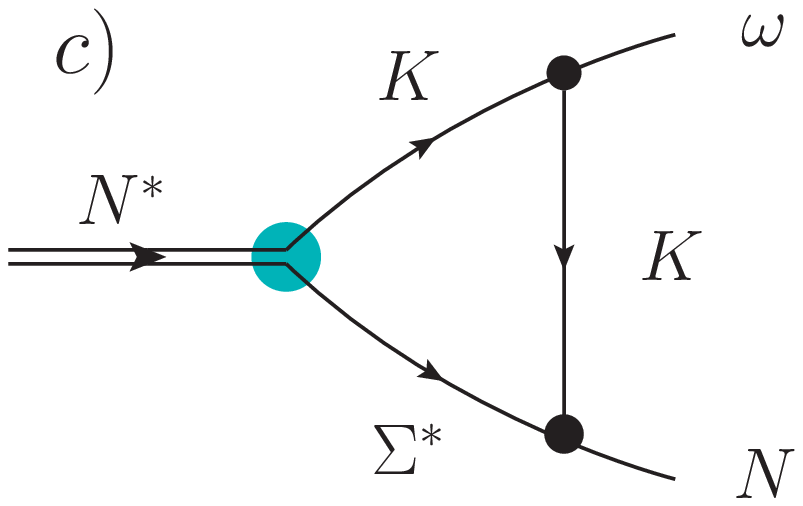}
\includegraphics[width=0.21\textwidth]{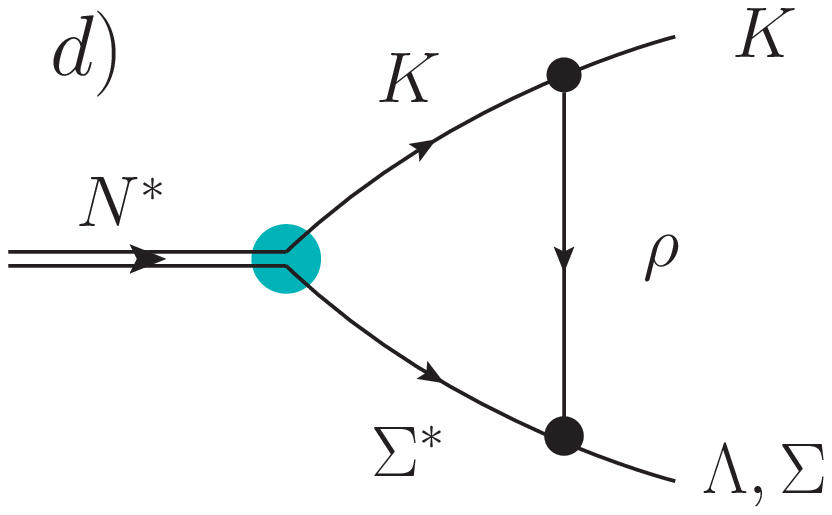}\\ \ \\
\includegraphics[width=0.21\textwidth]{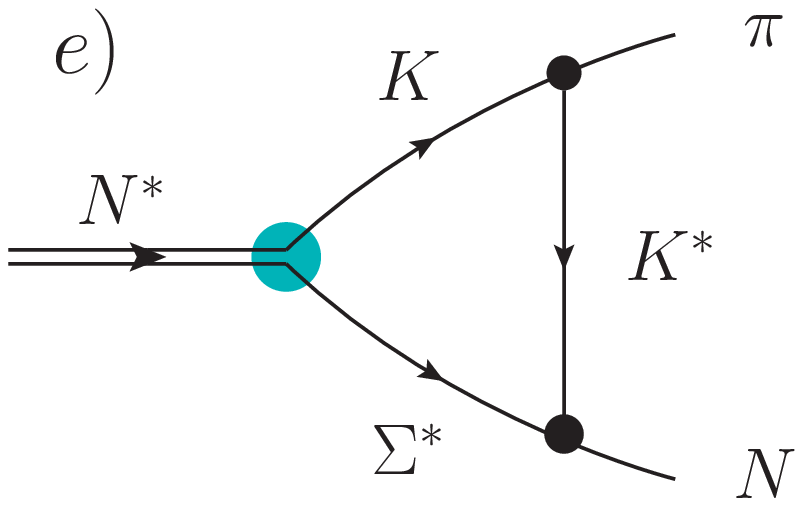}
\includegraphics[width=0.21\textwidth]{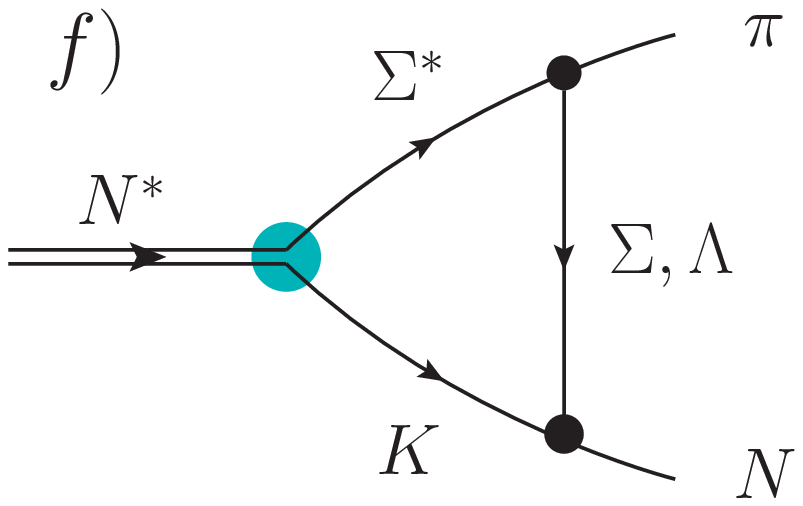}
\caption{(Color online) The decays of the $\Sigma^* K$ bound state. a)
$N\sigma$ channel with $K$ exchange. b) $N\rho$ channel
with $K$ exchange.  c) $N\omega$ channel
with $K$ exchange. d) $K\Lambda/\Sigma$ channel with $\rho$ exchange.
e) and f)  are for $N\pi$ decay channel with $K^*$
exchange and with $\Lambda/\Sigma$ exchange, respectively. \label{Fig: mechanism}}
\end{center}
\end{figure}

Since two-body decay of a molecular state occurs only through hadron loop mechanism, it is suggested that  three-body decay may be larger than two-body decay~\cite{He:2011ed}. As shown in Refs.~\cite{He:2011ed,He:2012zd,He:2013oma}, it was found that three-body decay has positive correlation to the decay width of the constituents. Hence, the three-body decay of the bound state $\Sigma^*K$ is suppressed due to the small decay width of $\Sigma^*$.
In this work the two-body decays through exchanging a particle between two constituents as shown in Fig.~\ref{Fig: mechanism} are taken as the main decay channels of the $\Sigma^*K$  bound state.

The decay amplitudes can be written as
\begin{eqnarray}
	{\cal M}&=&\sum_\lambda A_{\lambda}G_0|\Gamma_{\lambda}\rangle
	=\sum_\lambda A_{\lambda} FN^{-1}|\phi_\lambda\rangle
	\nonumber\\
	&\equiv&\sum_\lambda \int \frac{d^4
	k~\delta(k^0_1-E_1)}{(2\pi)^4}\frac{F(|{\bm k}|)}{N(|{\bm
	k}|}\phi_\lambda({\bm
	k}) A_{\lambda,\lambda'_1\lambda'_2}( {\bm k},{\bm k}'),\quad\label{Eq: amp}
\end{eqnarray}
where $\lambda'_{1,2}$ are helicities for two final particles
and ${\bm k}$ and ${\bm k}'$ are the momenta for $\Sigma^*$ and final meson in the center of mass frame.
$A_{\lambda,\lambda'\lambda_2'}$ is the amplitudes for two
constituents $\Sigma^*$ and $K$ to two final  particles, $N\pi$,
$N\sigma$ and so on. The  definitions of wave
function $\phi$ and $G_0$ have been used in the derivation of Eq.~(\ref{Eq: amp}).
The normalization of wave function  $\phi$ insures that there is no free
total factor in our calculation of amplitude.

Besides the Lagrangians in Eq.~(\ref{Eq: Lp}), the following Lagrangians are used to
calculate the amplitudes $A_{\lambda,\lambda'\lambda_2'}$,
\begin{eqnarray}
	{\cal L}_{KK\sigma}&=&g_{KK\sigma}{2m_\pi} \partial_\mu K^\dag
	\partial^\mu K\sigma,\\
	{\cal L}_{K^*K\pi}&=&ig_{K^*K\pi} K^{*\mu}({
	\pi}\partial^\mu K-\partial^\mu {\pi} K),\\
	{\cal
	L}_{KNY}&=&\frac{f_{KNY}}{m_N+m_Y}
	\bar{N}\gamma^\mu\gamma_5Y\partial_\mu K+H.c.,\\
		{\cal
	L}_{PB\Sigma^*}&=&\frac{f_{PB\Sigma^*}}{m_P}\partial_\mu
	{K}\bar{\Sigma}^{*\mu}N+H.c.,\\
	{\cal L}_{VB \Sigma^*}&=&-i\frac{f_{V
	B\Sigma^*}}{m_V}\bar{\Sigma}^{*\mu}\gamma^\nu\gamma_5
	[\partial^\mu{\rho}_\nu-\partial_\nu{
	\rho}^\mu]B	+H.c.,
\end{eqnarray}
where $PB$ means $KN$, $\pi\Lambda$ or
$\pi\Sigma$, $VB$ means $\rho\Lambda$, $\rho\Sigma$ or
$K^*N$ and $Y$ means $\Sigma$ or $\Lambda$.
The coupling constants are adopted as
$g^2_{KK\sigma}/4\pi=0.25$~\cite{Krehl:1997kg},  $g_{K^*K\pi}=-3.23$,
$f_{KN\Lambda}=13.24$ and $f_{KN\Sigma}=3.58$~\cite{Gao:2010ve}.
The coupling constants about $\Sigma^*$ can be obtained through $f_{\Sigma^*\Lambda\pi}=1.27$,
$f_{KN\Sigma^*}=-3.22$~\cite{Gao:2010ve} and $f_{\Delta\rho N}=-6.08$~\cite{Matsuyama:2006rp} with the $SU$(3) symmetry relations ${f_{\pi \Sigma\Sigma^*}}/{m_\pi}=-\frac{1}{\sqrt{3}}\frac{f_{\pi
\Lambda\Sigma^*}}{m_\pi}$,
$\frac{f_{\rho
\Lambda\Sigma^*}}{m_\rho}=\frac{1}{\sqrt{2}}f_{\Delta\rho N}/m_\rho$,
$\frac{f_{\rho
\Sigma\Sigma^*}}{m_\rho}=-\frac{1}{\sqrt{6}}f_{\Delta\rho N}/m_\rho$ and
$\frac{f_{K^* N\Sigma^*}}{m_{K^*}}=-\frac{1}{\sqrt{6}}f_{\Delta\rho N}/m_\rho$.

The $\Sigma^*$ carries spin-parity $J^P=3/2^+$ and isospin $I=1$. A system composed of $\Sigma^*$ and kaon  carries $I=1/2$ or $3/2$.
In this work, all states with $J\leq3/2$
are considered and the ranges of the cutoffs in form factors are chosen as  $1<\Lambda<5$ GeV.  The bound state solutions with the binding energies $E=m_1+m_2-W$ are listed in Table~\ref{diagrams}
and compared with the values from the PDG and the BnGa groups~\cite{Agashe:2014kda,Anisovich:2011fc}.
\renewcommand\tabcolsep{0.12cm}
\renewcommand{\arraystretch}{1.}
\begin{table}[hbtp!]
\caption{The binding energies $E$ for $\Sigma^*K$ system with different cut off $\Lambda$
	The cut off
	$\Lambda$, binding energy and branch ratio are in the units of GeV,	MeV and \%, respectively.
\label{diagrams}}
\begin{center}
\begin{tabular}{c|rr|rrrrrr}\bottomrule[1.pt]
$\Lambda$ & $E$ &  $\Gamma$ & $N\sigma$ &  $N\rho$ &
$N\omega$ &  $N\pi$ & $\Lambda K$ & $\Sigma K$\\\hline
  1.68   &   3  &   41&  55.9  &  4.7  &   14.1   &   22.4   &    2.3   &  0.6  \\
  1.72   &   8  &   73&  55.8  &  4.7  &   14.0   &   22.6   &    2.3   &  0.6  \\
  1.76   &  16  &  111&  55.7  &  4.7  &   14.0   &   22.7   &    2.2   &  0.6  \\
  1.80   &  28  &  155&  55.6  &  4.8  &   14.2   &   22.8   &    2.1   &  0.5  \\
  1.84   &  44  &  204&  55.3  &  4.9  &   14.6   &   22.7   &    2.0   &  0.5  \\
  1.88   &  67  &  257&  54.9  &  5.1  &   14.9   &   22.9   &    1.8   &  0.4  \\
  1.92   &  100 &  312&  53.6  &  5.1  &   14.7   &   24.8   &    1.5   &  0.3  \\\hline
  PDG \cite{Agashe:2014kda} &  $30^{+25}_{-25}$&        & $24^{+24}_{-24}$  &     $6^{+6}_{-6}$    &
  $20^{+4}_{-4}$  &   $7^{+6}_{-6}$   &    & $0.7^{+0.4}_{-0.4}$       \\
  BnGa \cite{Anisovich:2011fc}  &  $0^{+20}_{-20}$&    $200^{+20}_{-20}$    & $60^{+12}_{-12}$  &      &
  &   $3^{+2}_{-2}$   &   $4^{+2}_{-2}$ & $15^{+8}_{-8}$
  \\\hline
  $[N(\thalf^-)]_3$ & -85 &   $324$  & & $57.1$  &   $12.3$   &
   $20.8$   &   $9.7$ & $0$       \\
\bottomrule[1.pt]
\end{tabular}
\end{center}
\end{table}

Only one bound state solution with $I=1/2$ and $J^P=3/2^-$ is  found from the interaction of $\Sigma^*$ and $K$.
The decay width becomes larger with increase of the binding energy. It is understandable because the large binding energy means that the distance between two constituents is smaller so that the quark exchange is prone to happen in the bound state.  Compared with the PDG and BnGa values about mass and total width, the best cut off $\Lambda\approx1.80$ GeV is reasonable and consentient to the value in the
literature~\cite{Doring:2010ap}. The branch ratios of $N(1875)$
are stable compared with binding energy. The $N\sigma$ channel is the most
important decay channel, about $55\%$, which is consistent with the
PDG suggested values $24\pm24\%$ and $60\pm12\%$ from the BnGa analysis.
The main decay channel of the $[N3/2^-]_3$ predicted in the constituent  quark model is
$N\rho$ which is much larger than other decay channels. It
conflict with both the values suggested by the PDG and these obtained by the BnGa analysis.
Hence, the decay pattern of $N(1875)$
disfavors the assignment as $[N3/2^-]_3$.

\section{Summary}

In this work,the internal structures of the $(1875)$ and the $N(2120)$ are investigated. The experimental data for the photoproduction of $\Lambda(1520)$ off proton released by the CLAS and LEPS Collaborations suggest the explanation of the $N(2120)$ as the third state with $J^P=3/2^-$ in the constituent quark model. The $N(1875)$ is explained as a bound
state from the interaction of $\Sigma^*$ and kaon, which is supported by the numerical results of both binding energy and decay pattern of
the bound state of $\Sigma^* K$ system with isospin $I=1/2$ and spin-parity $J^P=3/2^-$.

\section*{Acknowledgments}
This project is partially supported by the Major State
Basic Research Development Program in China (No. 2014CB845405),
the National Natural Science
Foundation of China (Grants No. 11275235, No. 11035006)
and the Chinese Academy of Sciences (the Knowledge Innovation
Project under Grant No. KJCX2-EW-N01).


\end{document}